# X-ray fluorescence computed tomography (XFCT) imaging with a superfine pencil beam x-ray source

Ignacio O. Romero, Yile Fang, Michael Lun, and Changqing Li*
Department of Bioengineering, University of California, Merced, Merced, CA, USA.

## ABSTRACT

X-ray fluorescence computed tomography (XFCT) is a molecular imaging technique of x-ray photons, which can be used to sense different elements or nanoparticle (NP) agents inside deep samples or tissues. XFCT has been an active research topic for many years. However, XFCT has not been a popular molecular imaging tool because it has limited molecular sensitivity and spatial resolution. To further investigate XFCT imaging, we present a benchtop XFCT imaging system, in in which a unique pencil beam x-ray source and a ring of x-ray spectrometers were simulated using GATE (Geant4 Application for Tomographic Emission) software. An accelerated majorization minimization (MM) algorithm with an $L^1$ regularization scheme was used to reconstruct the XRF image of Molybdenum (Mo) NP targets from the numerical measurements of GATE simulations. With a low x-ray source output rate, good target localization was achieved with a DICE coefficient of 83.681%. The reconstructed signal intensity of the targets was found to be relatively proportional to the target concentrations if detector number and placement is optimized. The MM algorithm performance was compared with maximum likelihood expectation maximization (ML-EM) and filtered back projection (FBP) algorithms. In the future, the benchtop XFCT imaging system will be tested experimentally.

## 1. INTRODUCTION

X-ray fluorescence computed tomography (XFCT) is a molecular imaging technique of x-ray photons, which can be used to sense different elements or nanoparticle agents inside deep samples or tissues. XFCT has been an active research topic for many years. XFCT imaging quantifies and maps the distribution of a high atomic number (Z) element of interest in objects. Many XFCT benchtop systems employ a cone beam source geometry, with pinhole detector collimation to reduce the imaging time and dose [1-3]. However, the pencil beam geometry provides greater spatial resolution due to the radiation of a line rather than a volume at the cost of a longer scan time. Upon x-ray excitation, element specific characteristic x-rays are emitted from a target and then recorded. Ideally, multiple spectral detectors are configured to optimize the detected x-ray fluorescent signal and reduce the scan time and dose delivered to the imaging object [4].

Nanoparticles like Molybdenum (Mo) and Gold (Au) nanoparticles (NPs) have attracted significant attention in biomedical imaging. The K-shell emission energies of these NPs have greater penetrability which enables deeper functional tissue imaging. The high biocompatibility of MoNPs and AuNPs allows for greater injection doses with less concerns of cell toxicity which makes it feasible for the NPs to act as both CT contrast agents and functional imaging contrast agents [5,6]. For many cancer imaging applications, the NPs are used as passive targeting agents due to the enhanced permeability and retention (EPR) effects of the tumor [6]. AuNPs have been extensively investigated due to their high affinity ligands which have led to dose enhancements in radiation cancer treatment [3, 6-8].

A simple filtered back projection (FBP) reconstruction can be performed to obtain the x-ray fluorescent (XRF) image, however, the FBP is ill posed and small perturbations in the image data from scatter noise will cause significant deviations from the ground truth in the reconstructed image. A popular iterative method to reconstruct the emission tomographic image is the maximum likelihood expectation maximization (ML-EM) algorithm [1-4,9]. Recently, a Nesterov accelerated MM algorithm with an $L^1$ regularization known as fNUMOS (fast NonUniform Multiplicative MM algorithm with Ordered Subsets acceleration) has shown success in XLCT image reconstruction [10,11]. The MM algorithm with the Nesterov acceleration technique guarantees monotonicity and improves the convergence rate by limiting computational exhaustive matrix operations on the system matrix while promoting sparsity with the $L^1$ regularization [12,13].

In this work, a benchtop XFCT imaging system is presented in which a quasi-monochromatic pencil beam x-ray source from Sigray Inc and a ring of spectrometers were simulated using GATE (Geant4 Application for Tomographic Emission) [14]. A quasi-monochromatic x-ray source spectrum was simulated due to the emergence of compact quasi-monochromatic sources in laboratory settings which enhance CT image quality [15]. The x-ray propagation inside the media was modeled to construct the system matrix. The XRF image of two MoNP targets inside a cylindrical water phantom was reconstructed. The number of detectors used for the reconstruction was varied to show the detector number and position dependence on the reconstruction image quality. The images from the simulations and experimental were reconstructed using the fNUMOS algorithm and compared with the ML-EM and FBP algorithms.

This paper is organized as follows. In section 2, the methods of the GATE simulation using the unique Sigray source and ring detector configuration are presented. In section 3, the results showing the effects of detector number and placement, and the results comparing the fNUMOS reconstruction algorithm to ML-EM and FBP reconstructions are presented. The paper concludes with a discussion of the results and future works.

*cli32@ucmerced.edu;    http://biomedimaging.ucmerced.edu/home

## 2. METHODS

### 2.1 GATE simulation

The GATE software is a GEANT4 wrapper which utilizes the macro language to ease the learning curve of GEANT4 and allow GEANT4 to be more accessible to researchers [14]. The GATE simulations in this work were parallelized and executed with a custom bash script on a 20 CPU workstation. The simulation wait time was approximately two days. The physics lists enabled in GATE consisted of the photoelectric effect, Compton scattering, and Rayleigh scattering which are the primary physics processes accounted for in XFCT imaging. To observe characteristic x-rays, atom de-excitation was enabled under the photoelectric effect process. The GATE software stores all output as ROOT files [16]. The necessary data from the ROOT output file was extracted using custom C++ code and processed in MATLAB. The extracted data consisted of the detector element number, the deposited energy and the angular projection number. The number of data sets needed were equal to the number of linear translations.

A schematic of the GATE simulation setup along with a GATE simulation snapshot is seen in Figure 1. For demonstration purposes, only 10 x-ray photons are shown in the GATE simulation snapshot. A 30 mm diameter cylindrical water phantom was positioned at the center of the reference frame. Two targets (T1, T2) with uniform 5 mg/ml and 10 mg/ml concentrations of MoNPs were embedded offset from the phantom center with the 5 mg/ml target (T1) closest to the phantom surface.

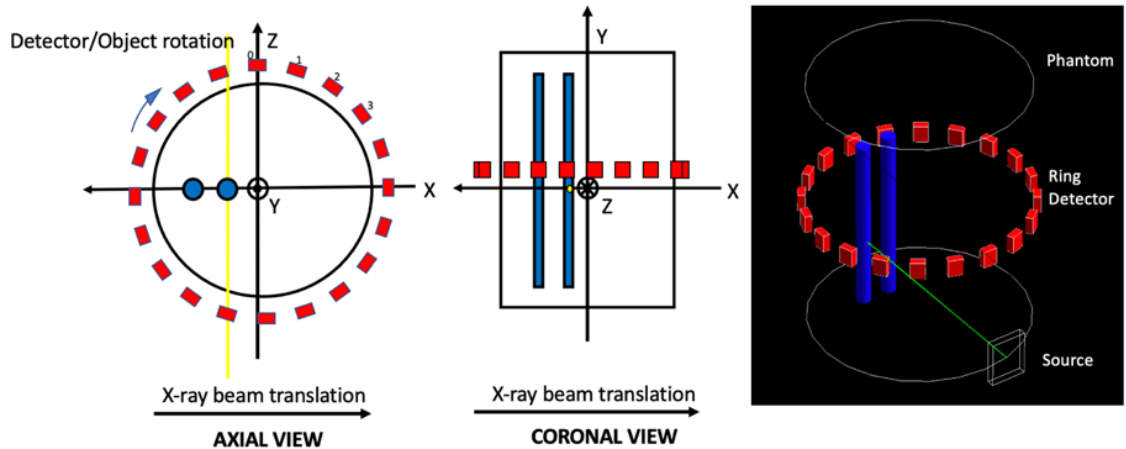

Figure 1: Axial view, coronal view, and snapshot of the XFCT simulation setup in GATE. The snapshot shows the trajectory of the x-ray photons as green lines.

The detector ring had a diameter of 31 mm and consisted of 20 Cadmium Zinc Telluride (CZT) elements. A detailed schematic of the ring detector is seen in Figure 2. Each detector element had dimensions of 2x2x1 $mm^3$. The x-ray beam size along the x-axis was 100 µm. $10^7$ x-ray photons per step were initialized. The x-ray beam scanned 2 mm below the detector ring. The linear step size was 125 µm therefore 248 linear steps were acquired to cover the phantom diameter and phantom edge. Six angular projections were acquired with 30° angular step size. In this setup, the detector ring was allowed to rotate with the phantom about the center of the reference frame while the x-ray beam translated linearly only.

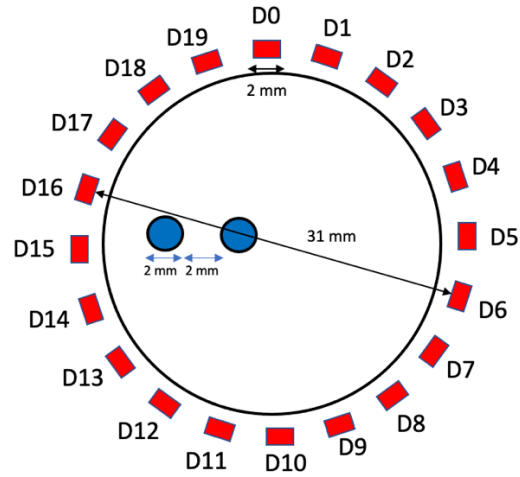

Figure 2: Schematic of the Cadmium Zinc Telluride (CZT) ring detector.

Due to the emergence of compact bright x-ray sources, the object was scanned by a modeled Sigray source. The Sigray source was simulated using the linear interpolation user spectrum tool from GATE. The energy of the emitted photon is determined according to a probability distribution created by piecewise-linear interpolation between the energies provided. The modeled Sigray source spectrum is seen in Figure 3.

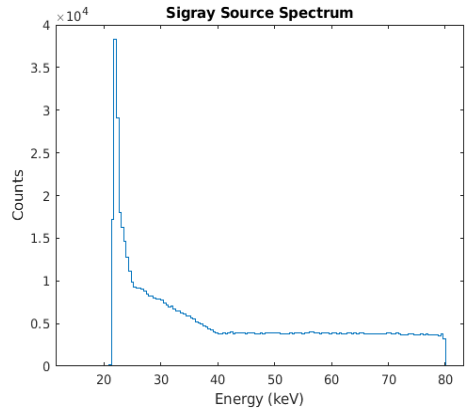

Figure 3: X-ray spectrum of a typical x-ray source from Sigray, Inc.

Before generating the x-ray fluorescent sinogram for reconstruction, scatter correction was performed. First an energy window was applied which was centered on the brightest fluorescent peak of the target element which was 17.48 keV. Then a 4$^{th}$ order polynomial interpolation was fitted on the acquired raw signal per linear step. The scatter signal obtained from the fitted line was removed from the fluorescent signal via subtraction. The energy resolution of the spectrometer used for the experimental component was considered when the net counts were collected. An example of the scatter correction method used for the acquisition is seen in Figure 4a. Only the counts bounded by the red vertical lines were collected. The net counts curve was shifted downward to show the effects of the correction, however, due to the unique spectrum of the Sigray source, little to no scatter is observed in the energy window. The corrected sinogram is seen in Figure 4b which shows a clean background.

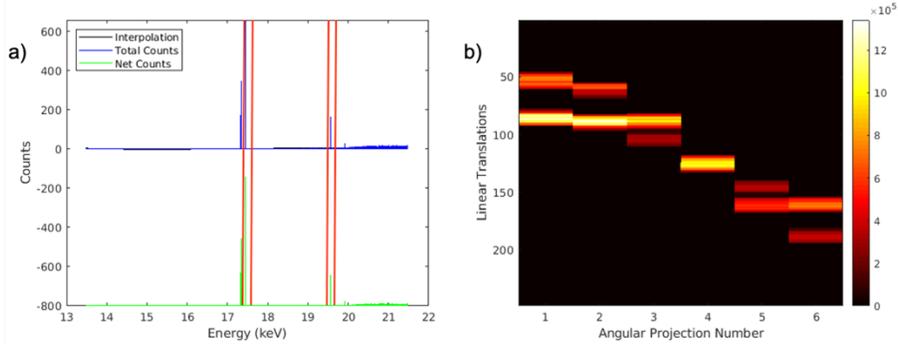

Figure 4: a) A 4th order polynomial interpolation fit for the removal of scattered x-ray photons. b) Corrected sinogram summed over all 20 detectors.

## 2.2 System matrix and reconstruction algorithms

To generate the system matrix, the physical process and the imaging geometric parameters were considered. In XFCT imaging, while the x-ray beam scans the object along a straight line, the x-ray beam intensity distribution along the beam line follows the Beer-Lambert law. Upon x-ray beam excitation, target nanoparticles emit isotropic fluorescent x-ray lines. The intensity distribution of fluorescent x-ray lines that span the detector surface similarly follow the Beer-Lambert law. The system matrix is composed of these excitation and sensitivity matrices, (**F** and **P**, respectively) as follows:

$$F_{j,m} = \exp\left(\sum -\mu_e(\mathbf{r}) \times L(\mathbf{r})\right) \quad (1)$$

$$P_{i,m} = \sum_{1}^{d_n} \exp\left(\sum -\mu_f(\mathbf{r}) \times L(\mathbf{r})\right) \quad (2)$$

$$\mathbf{A}_{n_d \times I \times J, m} = \begin{bmatrix} \begin{bmatrix} P_{1,m} \\ \vdots \\ P_{n_d,m} \end{bmatrix} \otimes F_{1,m} \\ \vdots \\ \begin{bmatrix} P_{1,m} \\ \vdots \\ P_{n_d,m} \end{bmatrix} \otimes F_{I \times J, m} \end{bmatrix} \quad (3)$$

$$0 \leq \mathbf{A}_{n_d \times I \times J, m} \leq 1$$

$\mu_e(\mathbf{r})$ is the linear attenuation coefficients of the imaging object and targets at the excitation energy. $\mu_f(\mathbf{r})$ is the linear attenuation coefficients of the imaging object and targets at the fluorescent energy. $L(\mathbf{r})$ is the distance from the start of the x-ray beam or fluorescent emission to the position **r**. $d_n$ is the discretization number of the detector surface for which the fluorescent x-ray lines will be accounted for. $\otimes$ represents the tensor product between $P_{i,m}$ and $F_{j,m}$ where $i \in [1, n_d]$ is the detector number, $j \in [1, I \times J]$ is the excitation scan number and m is the number of pixels used to discretize the object space. $I \times J$ is the product between angular projections number (I) and linear translation steps (J) of each angular projection, respectively. The forward model then becomes:

$$\mathbf{A}_{n_d \times I \times J, m} \mathbf{X}_{m,1} = \mathbf{B}_{n_d \times I \times J, 1} \quad (4)$$

where $\mathbf{X}_{m,1}$ is the unknown image vector to reconstruct and $\mathbf{B}_{n_d \times I \times J, 1}$ is the set of measurements.

The XFCT reconstruction can then be solved by minimizing the following optimization problem with the nonnegativity constraint.

$$\widehat{\mathbf{X}} = \arg\min_{\mathbf{x} \geq 0} \mathbf{Q}(\mathbf{x}) := \frac{1}{2} \|\mathbf{B} - \mathbf{AX}\|_2^2 + \lambda \|\mathbf{X}\|_1^1 \tag{5}$$

where $\lambda$ is the regularization parameter and $\|\mathbf{X}\|_1^1$ is the $L^1$ norm of the image vector $\mathbf{X}$. The fNUMOS algorithm is applied to minimize the $L^1$ regularized difference between the measurements modeled in GATE and the system matrix estimates. The details of the fNUMOS algorithm are explained in detail in [12]. The $L^1$ regularization term is employed since it is well known for sparsity enhancement [11]. The system matrix generation and image reconstruction using fNUMOS was performed in MATLAB.

The FBP and ML-EM image reconstructions were performed using the Michigan image reconstruction toolbox (MIRT) in MATLAB [17].

### 2.3 Evaluation Criteria

Two criteria were used to evaluate the quality of the reconstructed images as described in ref [9].

Contrast to noise ratio (CNR) measurements are usually performed to evaluate the quality of the reconstructed target signals. However, due to the zero background of the reconstructed images, only the contrast signal of each target was measured by taking the mean of the reconstructed signal of a 5×5 pixel region of interest taken from the target center. A target contrast ratio was calculated to determine the proportionality of the contrast signals between the targets by taking the ratio of the contrast of the 10 mg/ml concentration target (T2) to the 5 mg/ml concentration target (T1). For this simulation, the closer the target contrast ratio is to two the better.

The Dice similarity coefficient (DICE) was used to measure the accuracy of the reconstructed target localization by comparing the reconstructed and true target positioning.

$$\text{DICE} = \frac{2 \times |\text{ROI}_r \cap \text{ROI}_t|}{|\text{ROI}_r| + |\text{ROI}_t|} \times 100\% \tag{6}$$

where $\text{ROI}_r$ is the reconstructed target region of interest. $\text{ROI}_t$ is the true target region of interest. Generally, the closer DICE is to 100%, the better.

## 3. RESULTS

### 3.2 Effects of detector number and detector placement

To explore how the number of detectors and their placement along the ring configuration influence the quality of the reconstructed images, the number of detectors used for reconstruction were varied. The results of reconstructed images for varying detector number are seen in Figure 5. Zoomed in target regions are provided where the green circles represent the true target size and locations. From the dashed blue lines in the zoomed in images, line profiles are plotted with the intensity values normalized with respect to the max value along the dashed blue line. All detector number variations were able to completely separate and localize the target signals within the true target size without noise artifacts. All detector number variations were able to reconstruct uniform target signals, except for the 2-detector configuration.

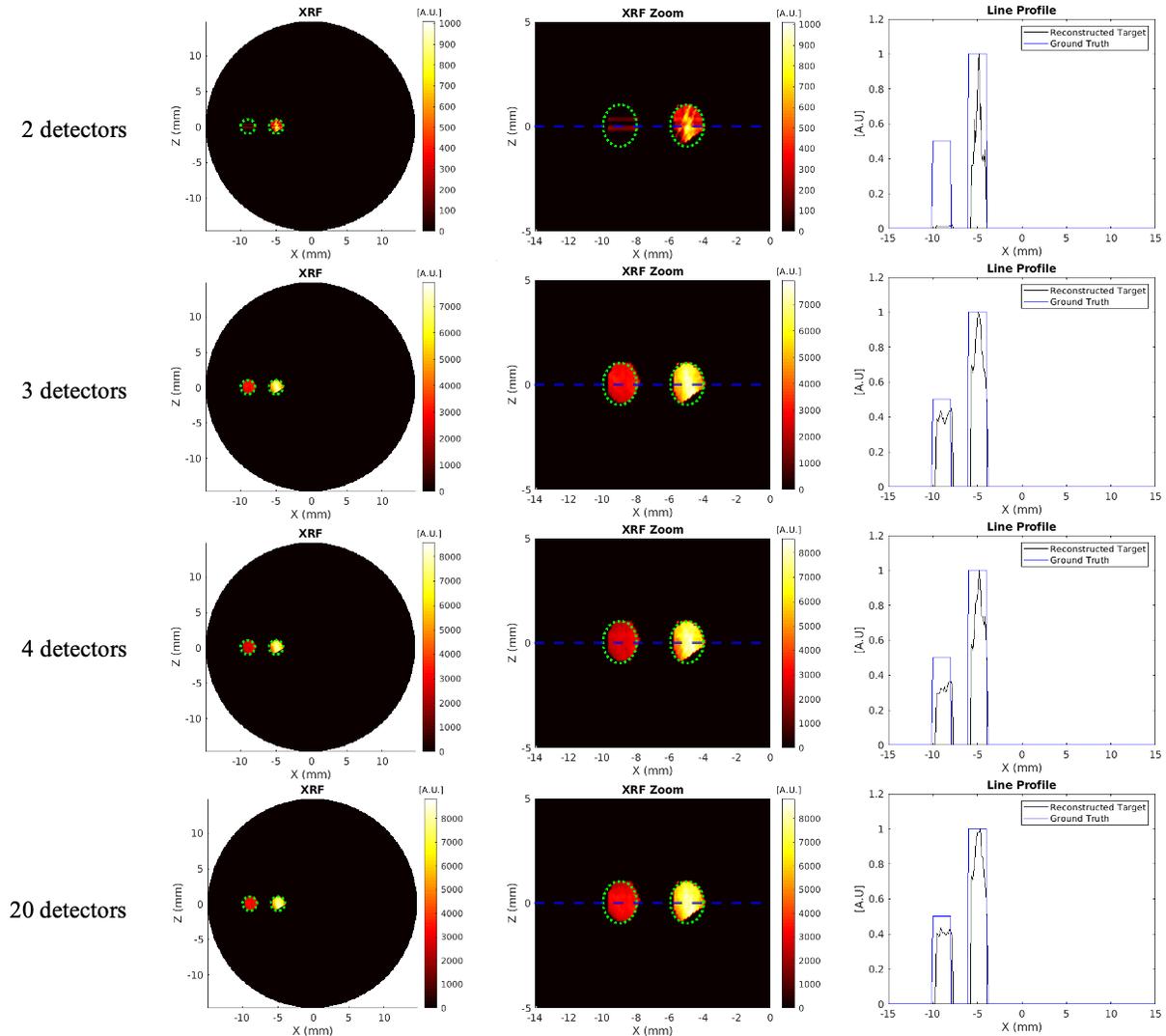

Figure 5: fNUMOS image reconstruction using different detector numbers. The full image (left) zoomed in target regions (middle), and line profiles (right) are shown. The green dotted line in the zoomed in target region indicates the exact target size and position, the blue dotted line indicates the line profile location.

The image quality metrics for the reconstruction with 2, 3, 4, and 20 detectors at different positions are listed in Table 1. All detector positions with the exception of the 2-detector configuration using the D0 and D4 detectors achieved a DICE coefficient of 83.681%. When the 2-detector configuration is changed to the D9 and D14 detectors, the DICE improves from 67.769% to 83.681%. The target contrast ratio also improved from 7.532 to 2.47 when using the D9 and D14 detector configuration. For the 3-detector configuration, the detectors D3, D12, and D14 provided a more accurate target contrast ratio than the D0, D9 and D14 detectors. For the 4-detector configuration, the D3, D8, D13, and D14 detectors provided a more accurate target contrast ratio than the D0, D5, D10, and D15 detectors. The improvements in the image quality metrics between varying detector positions highlights the importance of detector position in XFCT imaging.

Table 1: Image quality metrics for the GATE simulations with varying ring detector element number and position.

| Detector Number | Target Contrast Ratio | DICE (%) |
| --- | --- | --- |
| D0, D4 | 7.533 | 67.769 |
| D9, D14 | 2.470 | 83.681 |
| D0, D9, D14 | 2.489 | 83.681 |
| D3, D12, D17 | 2.026 | 83.681 |
| D0, D5, D10, D15 | 2.841 | 83.681 |
| D3, D8, D13, D18 | 1.435 | 83.681 |
| D0 to D19 | 2.489 | 83.681 |

### 3.2 Reconstruction Algorithm Comparison

To explore the quality of the fNUMOS image reconstruction, the fNUMOS image reconstruction was compared to the ML-EM and FBP image reconstruction. The results were based on the 20-detector configuration. For the ML-EM and FBP algorithm, a 50% max threshold was implemented to remove noise artifacts. A Hann filter was implemented in the FBP algorithm.

Figure 6 shows the reconstructed images and target line profiles of the fNUMOS, ML-EM, and FBP algorithms. In general, the fNUMOS algorithm was superior in terms of achieving a uniform target signal reconstruction with no background noise and target signal which were proportional to their respective concentrations of 5 mg/ml and 10 mg/ml. Neither the ML-EM nor FBP algorithms were able to reconstruct the target signals with intensities proportional to the target concentrations.

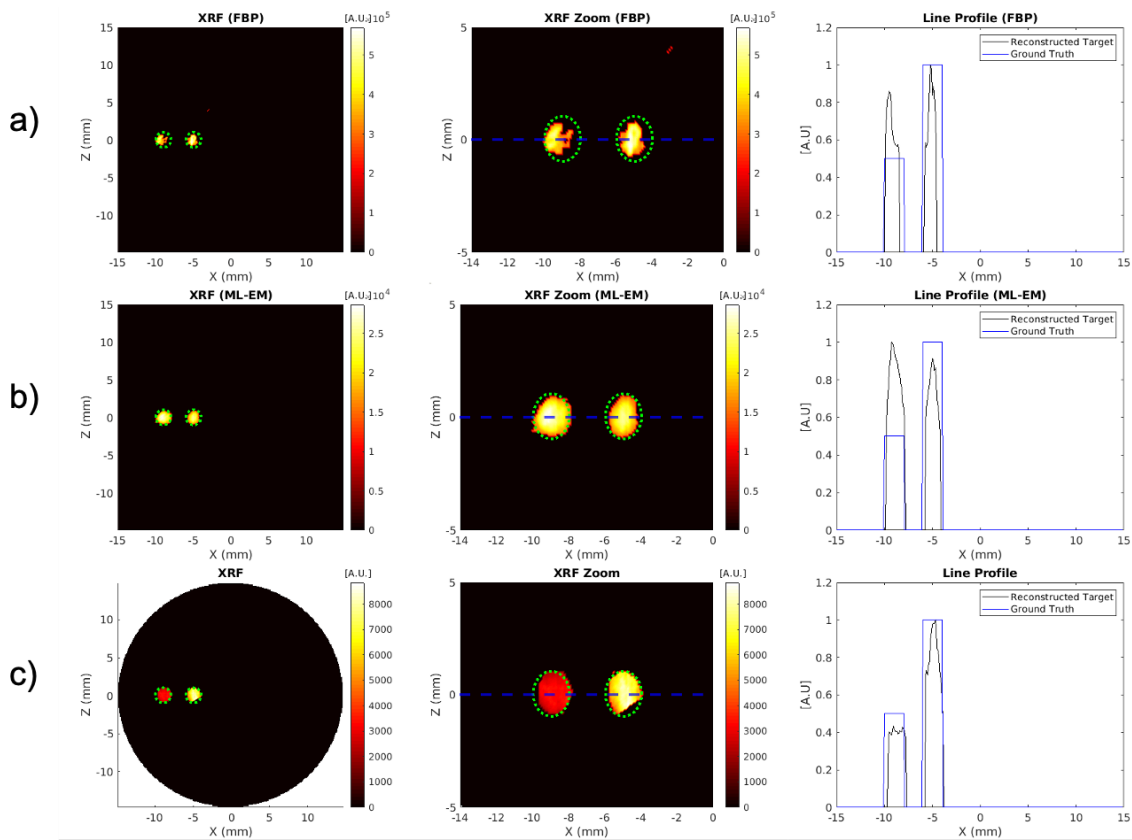

Figure 6: a) image reconstruction results using FBP. b) Image reconstruction results using ML-EM. c) Image reconstruction results using fNUMOS. All reconstruction algorithms utilized all 20 detectors from the ring detector configuration.

Table 2 shows the image quality metrics of reconstructed images from the fNUMOS, ML-EM, and FBP algorithms. The ML-EM algorithm gave superior contrast measurements and DICE coefficients than fNUMOS, but the target contrast ratio did not reflect the correct target concentration ratio. The FBP achieved a better target contrast ratio than ML-EM, but the DICE coefficient was significantly inferior to ML-EM and fNUMOS. The combination of the accuracy of the target contrast ratio and DICE coefficient shows that the fNUMOS outperformed the other algorithms.

It is also worth noting the time taken for the reconstruction algorithms to reach their solutions. The fNUMOS algorithm takes < 0.05 seconds to successfully reconstruct the 300 x 300 XRF image using the data collected by all 20 detectors in the ring configuration. The ML-EM algorithm consumes >60 seconds and the FBP algorithm consumes < 0.5 seconds to reconstruct the same image with overall image quality below that of fNUMOS reconstruction.

Table 2: Image quality metrics for the GATE simulations with FBP, ML-EM, and fNUMOS reconstruction.

| Reconstruction Method | Reconstruction Time (s) | Target Contrast Ratio | DICE (%) |
|---|---|---|---|
| FBP | 0.479048 | 2.410 | 64.918 |
| ML-EM | 61.89206 | 0.958 | 86.138 |
| fNUMOS | 0.015264 | 2.489 | 83.681 |

## 4. DISCUSSIONS AND CONCLUSIONS

In this work, a benchtop XFCT imaging system comprised of ring detector elements and a benchtop quasi-monochromatic source was modeled in GATE. The fNUMOS algorithm was applied to reconstruct the image successfully and efficiently. Due to the unique energy characteristics of the benchtop source, a relatively low x-ray photon number is enough to obtain a good XRF signal even when a low number of detectors from the ring configuration were used.

The proposed imaging scheme and reconstruction algorithm was studied with different detector number and positions. Compton scattering effects prevent a good and clear XRF signal from being detected. Detector placement is crucial to minimize the detection of greater scattering counts. Typically, detectors are placed orthogonal to the incident beam trajectory to minimize scatter. In this work, it was shown that 2 detectors are enough to accurately reconstruct the targets, however, the accuracy of the reconstruction is strongly dependent on the placement of the detectors as seen in Table 1. Placing detectors at approximately 180° and 90° from the incident beam trajectory is best for the two-detector case.

The fNUMOS reconstruction algorithm performance was compared with the popular ML-EM and FBP algorithms. The image quality metrics show that the fNUMOS reconstruction algorithm is superior when considering the combination of the target contrast ratio, the DICE coefficient measurements, and reconstruction time. One possible explanation is that fNUMOS is good at imaging sparse targets as seen in this work and XLCT studies [10, 11].

In summary, a benchtop XFCT imaging system with a ring detector and a unique quasi-monochromatic benchtop source was proposed, and GATE simulations were performed to validate its feasibility. This work provides further emphasis for the need of unique and bright x-ray sources like the Sigray source to become more available in the laboratory setting which will lead to reduced dose and scan time.

In the future design of the benchtop XFCT system, a compact quasi-monochromatic source will be used to improve the XRF signal for better XRF imaging results. The proposed x-ray source from Sigray Inc has unique properties that enable a higher heat load on the anode and reduce exposure switching times. We aim to collaborate with the company to improve of the benchtop XFCT design by increasing the sensitivity and reducing scanning times. The experimental XFCT results will be compared with the simulated XFCT results given similar imaging setup conditions.